\documentclass[twocolumn,showpacs,preprintnumbers,amsmath,amssymb]{revtex4}
\usepackage{epsfig,graphicx}
\usepackage{dcolumn}
\usepackage{bm}

\begin{document}
\draft

\title{Diversity of rationality affects cooperation in spatial prisoner's dilemma game}

\author{Yu-Zhong Chen\footnote{For correspondence: cyz001383@gmail.com}, Zi-Gang Huang, Sheng-Jun Wang, Yan Zhang and
Ying-Hai Wang\footnote{For correspondence: yhwang@lzu.edu.cn}}

\address{Institute of Theoretical Physics, Lanzhou University, Lanzhou Gansu 730000,
China}

\date\today

\begin{abstract}
In real world, individual rationality varies for the sake of the
diversity of people's individuality. In order to investigate how
diversity of agent's rationality affects the evolution of
cooperation, we introduce the individual rationality proportional to
the $\beta$th power of the each agent's degree. Simulation results
on heterogeneous scale-free network show that the dynamic process is
greatly affected by the diversity of rationality. Both promotion and
inhibition of cooperative behavior can be observed at different
region of parameter $\beta$. We present explanation to these results
by quantitative and qualitative analysis. The nodes with middle
degree value are found to play a critical role in the evolutionary
processes. The inspiration from our work may provide us a deeper
comprehension towards some social phenomenon.
\end{abstract}

\maketitle

\section{introduction}

In evolutionary biology, behavioral sciences, and more recently in
economics, understanding conditions for the emergence and
maintenance of cooperative behavior among selfish individuals
becomes a central issue \cite{vonNeumann,MaynardSmith}. Including
cooperation and defection as the two competing strategies, the
prisoner's dilemma game (PDG) is regarded as a paradigm for studying
this issue \cite{Axelrod}-\cite{M.M-Gibbons1997}.

Based on a structured population
\cite{NowakMayNature19921993}-\cite{LiebermanHauertNowakNature2005},
considerable efforts have been extended by allowing the players to
voluntary participating \cite{SzaboHauertPRL2002}, or introducing
dynamic network model \cite{ZimmermannPRE2004,huangzigangEPJB2007},
dynamic payoff matrices \cite{TomochiPRE2002}, dynamic preferential
selection \cite{wuzhixiPRE2006}, and difference between interaction
and learning neighborhoods \cite{wuzhixiPRE2007}. Santos and Pacheco
\cite{PDGscalefree} have studied the PDG on heterogeneous scale-free
networks, and observed that, when the underlying network is
scale-free, cooperation can be greatly enhanced and becomes the
dominating trait throughout the entire range of parameters of the
game, due to the cooperators' cluster forming nature
\cite{CclusterPRL2007}.

In the mentioned works, individual particularity is not the main
topic. However, particularity is ubiquitous among the individuals of
social groups and animal species. Thus, the diversity of
individuality inevitably appears between the players engaging in the
evolutionary games. Instead of taking individual difference into
account directly, some works concentrate on the individual
similarity
\cite{RioloCohenAxelrodNature2001}-\cite{TraulsenClaussenPRE2004}.
Recently, it has been directly proved that diversity of certain
individual property can efficiently promote cooperative behavior in
evolutionary games \cite{socialdiversity,socialdiversityNature}. The
authors introduce social member's extrinsically determined
properties, like wealth or social status, to increase or decrease
the fitness of a player depending on its location on the spatial
grid. Different scaling factors are provided to different nodes,
rescaling their payoff matrix in PDG \cite{socialdiversity}.

Different from the diversity of extrinsically determined properties
\cite{socialdiversity,socialdiversityNature}, this paper
concentrates on the diversity of the intrinsic property, individual
rationality. Szab\'{o}'s stochastic evolutionary rule
\cite{GSzabo1998}, especially the Fermi upgrading rule, has taken
this vital and intrinsically determined property into account. In
the formula of Fermi rule, the variable temperature indicates how
rational the individual is, when making decision in the game. Just
as temperature in statistical physics, this very variable, in former
works \cite{GSzabo20052006}-\cite{Jeromos2} was viewed mainly as a
stochastic noise. Phenomena like stochastic resonance
\cite{wwxnetworkrandomness} and second-order phase transition
\cite{Jeromos1} are discovered. These works have considered the
individual rationality to be at the same value for every player in
the game. Therefore, this variable actually serves as a reflection
of collective rationality belonged to the whole system. However, in
real society, individual rationality depends on its intelligence,
disposition, motivation, and circumstance, which differ from
individual to individual. Serving as an intrinsic factor, different
level of rationality determines different choice and correspondingly
fosters different game result. Within our study, we regard the
degree of a node as a rank of the individual's certain social
feature, for instance, social status, and directly relate this
feature to the individual rationality. An agent's rationality is set
to be proportional to the $\beta$th power of the agent's degree
\cite{bataformula}. In this way, the diversity of rationality is
associated with the diversity of degree. Networks with a
heterogeneous topological structure are used in our work. We
reported below that cooperation is enhanced in a certain region of
parameter $\beta$, but inhibited in other region.

The model and simulation result are presented in section II.  To
explore the mechanism for both promotion and inhibition of
cooperation, we present a statistical analysis to
the dynamic process in terms of microscopic arguments in section
III. Final conclusion is to be drawn in section IV, as well as some
sociological inspirations from our findings.

\section{the Model and Simulation Result}

To introduce the diversity of the intrinsic property, rationality,
we consider an evolutionary two-strategy prisoner's dilemma game
with players located on vertices of a heterogeneous network. Each
individual is allowed to interact with its nearest neighbors, and
self-interactions are excluded. Players can adopt one of the two
simplest strategies: ``cooperate'' ($C$) and ``defect'' ($D$). The
strategy adoption mechanism is based on the rescaled version of the
payoff matrix introduced by Nowak \cite{nowak2}:
\begin{equation}
\begin{array}{ccc}
\ & \begin{array}{cc} C & D \end{array} \\
\begin{array}{c} C \\ D \end{array}&\left(\begin{array}{cc} 1&0\\ b&0
\end{array}\right)
\end{array} (1<b<2)
\end{equation}
During the evolutionary process, a player located on node $i$ can
follow the strategy of one of its randomly chosen neighbor at node
$j$, with the probability depending on the payoff difference
$(M_{i}-M_{j})$,
\begin{equation}
W_{ij}=\frac{1}{1+exp[(M_{i}-M_{j})/T_{i} ]} \label{Fermi}
\end{equation}
This is the Fermi updating rule \cite{GSzabo1998}, where $T_{i}$
characterize the level of rationality pertained to node $i$. And
$T_{i}=0$ denotes complete rationality, where the individual always
adopts the best strategy determinately; while $T_{i}>0$, it
introduces some irrational factor, that there is small possibility
to select the worse one; $T_{i}\rightarrow \infty$ denotes that the
individual is completely irrational, and its decision is random.
Within this study we consider the diversity of rationality defined
by the following function \cite{bataformula}:
\begin{equation}
 T_{i}=NT_{0}\frac {k_{i}^{\beta}}{\sum_{l}k_{l}^{\beta}}\label{T}
\end{equation}
where $N$ is the total number of nodes in the network, and $k_i$ is
the degree of node $i$. Here we adopt the Barab\'{a}si-Albert
scale-free network \cite{barabasi1,barabasi2}. $T_{0}$ denotes the
average value of rationality. We use parameter $\beta$ to tune the
relationship between node $i$'s degree $k_i$ and its rationality
$T_i$. While $\beta<0$, nodes with larger degree gain lower value of
rationality. While $\beta=0$, rationality is uniformly distributed.
$\beta>0$ denotes a reversed situation as compared with the case
$\beta<0$: nodes with larger degree gain larger value of
rationality. If $\beta=1$, rationality is distributed by power-law.
It is worth mentioning that significantly high value of $T_i$ could
induce substantially random behavior of an agent, though payoff
difference may be large. On the contrary, very low $T_i$ would
heavily enhance agent's sensitivity towards higher payoff.

\begin{figure}
\centerline{\resizebox{11cm}{!}{\includegraphics{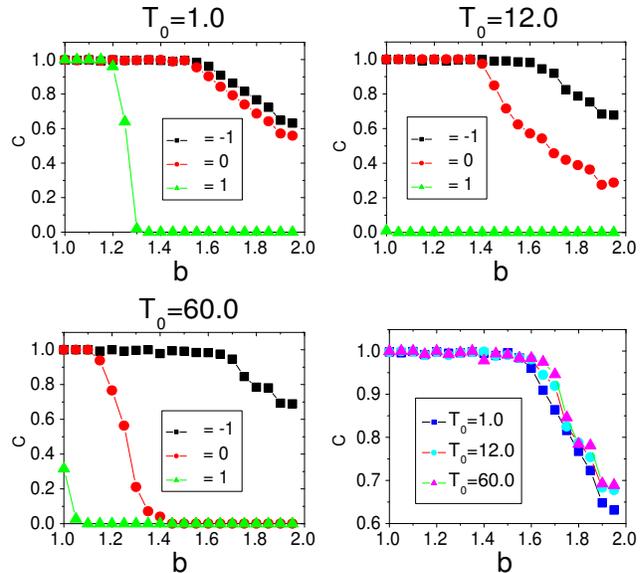}}}
\caption{The frequencies of cooperators $\rho_{c}$ versus the
temptation to defect $b$ at different average value of rationality
$T_0=1$, $12$, and $60$, and $\beta=-1$, $0$, and $1$.}
\label{figrouCb}
\end{figure}

\begin{figure}
\centerline{\resizebox{11.5cm}{!}{\includegraphics{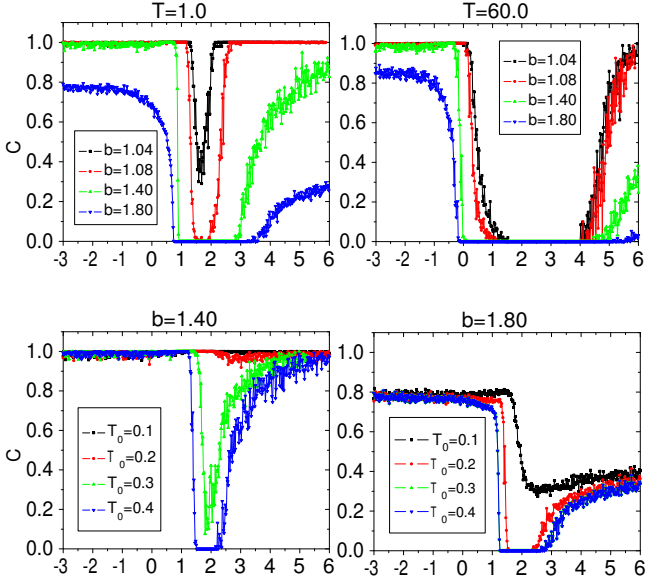}}}
\caption{ The frequencies of cooperators $¦Ñ_{C}$ versus $\beta$.
Left: The temptation to defect $b=1.04$, $1.08$, $1.6$, and $1.75$.
The average value of rationality $T_0=1$. Middle:  The temptation to
defect $b=1.4$. The average value of rationality $T_0=0.1$, $0.2$,
$0.3$, $0.4$, and $0.5$. Right:  The temptation to defect $b=1.8$.
The average value of rationality $T_0=0.1$, $0.2$, $0.3$, and
$0.4$.} \label{figrouCbata}
\end{figure}

The case of $\beta=0$ is discussed in
\cite{PDGscalefree,CclusterPRL2007}. Although the updating mechanism
we adopt is different from \cite{PDGscalefree,CclusterPRL2007}, when
$T$ is low, at a microscopic scale the following fact still exist:
cooperators tend to occupy the hubs, since hubs are directly
connected, if a defector occasionally takes over one hub, the
probability that it gets reoccupied by a cooperator becomes
essentially one. However, while using Fermi updating rule, this fact
may be affected by high rationality value. Fig.\ref{figrouCb} is
plotted to explore the influence of rationality in both cases,
$\beta=0$ and $\beta\ne0$. The BA network is built with the initial
number of nodes: $m_{0}=2$, the number of edges linked to the
exiting nodes from the newly added node in each time step: $m=2$,
and the average degree $\bar{k} =4$. The total number of nodes is
$N=4225$. Before the start of each game simulation, both strategies
populate the spatial grid uniformly. We adopted a synchronous
updating scheme. All the simulation results were obtained by
averaging over $1000$ generations after a transient time of $5000$
generations. Each data is obtained by averaging over ten different
network realizations with ten runs for each realization.

In Fig.\ref{figrouCb}, in the case of $\beta=0$, when $T_0=1$,
cooperation is dominating over the entire range of the temptation to
defect $b$, but sharp decrease of the frequencies of cooperators
$\rho_C$ can be measured when $T_0$ becomes large. This indicates
that the cooperation promoted by hubs is indeed sensitive to the
value of rationality. At the same time, in the case of $\beta=-1$,
the value of $\rho_C$ is not evidently affected by intense variation
of $T_0$, and there always exists a broad range in the parameter
space within which cooperation rule completely. Thus, there must be
other factors contributing to the facilitation of cooperation.

To further investigate how cooperation is influenced by the
parameter $\beta$, the variation of $\rho_C$ versus $\beta$ is
demonstrated in Fig.\ref{figrouCbata}. It is observed that
$\rho_{c}$ and $\beta$ have apparently non-monotonous relationship.
The shapes of the curves are similar to a gorge located in a
plateau. This shape indicates that diversity of rationality can
either promote or inhibit cooperation, depending on the value of
$\beta$. To highlight the sharp contrast between the effective
promotion and serious inhibition, in this paper the gorge is called
\emph{cooperation crisis}. The two downside graphes of
Fig.\ref{figrouCbata} show the cases of relatively small values of
average rationality, where the similar phenomenon is observed.
Higher value of $T_0$ or $b$ only results in larger width of the
gorge, leaving the shape of the curves unchanged. These results
reveal that the evolutionary dynamics is particularly sensitive to
rationality distribution. In order to explain the main features of
the reported results, especially the \emph{cooperation crisis}, we
hereafter scrutinize in depth the microscopic evolution of
cooperation.

\section{Analysis and discussion}

When $\beta=0$, in \cite{PDGscalefree,CclusterPRL2007}, the
prevalence of cooperation is because that on the heterogeneous
network topology, hubs can stick together the cooperator cycles that
would otherwise be disconnected, and form stable cooperative
clusters (C cluster) \cite{CclusterPRL2007}. In
\cite{CclusterPRL2007}, a cooperative cluster, namely a cooperator
core, is a connected component fully and permanently occupied by
pure cooperators. While invaded by defectors, the local structure of
a C cluster can be viewed as a C strategy hub surrounded by a number
of periphery neighbors, most of which are cooperators. Then whether
C clusters are stable or not when $\beta\neq 0$ is to be analyzed.
Two opposite effects generated by four crucial dynamic processes
determine the fluctuation number of cooperators and defectors in a C
cluster in every next time step, and thus determine the stability of
the C cluster:
\begin{description}
\item[Effect1:] Corruption of C cluster
     \begin{description}
     \item[Process (A):]the hub node of a C cluster adopts the strategy of a periphery defector, and then transits to D strategy;
     \item[Process (D):]a periphery cooperator adopts the strategy of the hub defector, and then transits to D strategy.
     \end{description}
\item[Effect2:] Consolidation of C cluster
     \begin{description}
     \item[Process (B):]a hub defector adopts the strategy of a periphery cooperator, and then transits to C strategy;
     \item[Process (C):]a periphery defector adopts the strategy of the hub node of a C cluster, and then transits to C strategy.
     \end{description}
\end{description}

The four processes are corresponding to four kinds of strategy
transition probability according to the Fermi updating rule. Based
on mean-field approximation, imaging a localized block in the
network, a node is surrounded by $k$ neighbors among which the
cooperators have a proportion of $\mu$ while the defectors have the
rest fraction $(1-\mu)$. The payoff difference between a cooperator
and a defector can be denoted by $\mu(k_{i}-k_{j}b)$ or
$\mu(k_{i}b-k_{j})$. Because the mean-field hypothesis is not always
fit for the evolutionary games on networks, the following analysis
can only be qualitative. From equation (\ref{Fermi}) and (\ref{T}),
we gain the four kinds of transition probability:
\begin{equation}
\textrm{Effect1}\left\{ \begin{array}{ll} \textrm{Process(A):}&W_{C
\rightarrow D}^{H\rightarrow
P}=\frac{1}{1+exp[\frac{\mu(k_{H}-k_{P}b)}{T_{0}} (\bar
{k^{\beta}}/k_{H}^{\beta})]}\\
\textrm{Process(D):}& W_{C \rightarrow D}^{P\rightarrow
H}=\frac{1}{1+exp[\frac{\mu(k_{P}-k_{H}b)}{T_{0}} (\bar
{k^{\beta}}/k_{P}^{\beta})]}
\end{array}\right.
\end{equation}
\begin{equation}
\textrm{Effect2}\left\{ \begin{array}{ll} \textrm{Process(B):}&W_{D
\rightarrow C}^{H\rightarrow
P}=\frac{1}{1+exp[\frac{\mu(k_{H}b-k_{P})}{T_{0}} (\bar
{k^{\beta}}/k_{H}^{\beta})]}\\
\textrm{Process(C):}& W_{D \rightarrow C}^{P\rightarrow
H}=\frac{1}{1+exp[\frac{\mu(k_{P}b-k_{H})}{T_{0}} (\bar
{k^{\beta}}/k_{P}^{\beta})]}
\end{array}\right.
\end{equation}
The upper scripts $H$ and $P$ denote hub and periphery respectively.
Through these two formulas, the term $(\bar {k^{\beta}}/k^{\beta})$
remodifies and extends the Fermi rule.

\begin{figure}
\centerline{\resizebox{11.5cm}{!}{\includegraphics{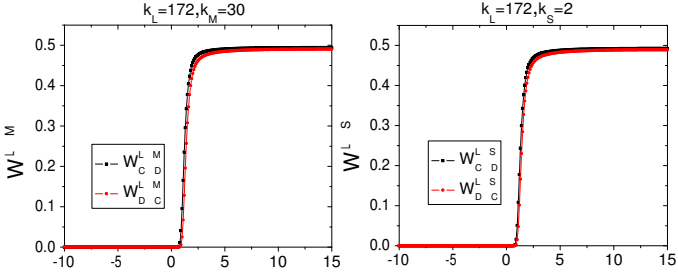}}}
\caption{The probability of high degree nodes' strategies transiting
to the strategies of middle (left) and small (right) degree nodes.
$k_L=172$, $k_M=30$, and $k_S=2$. The black lines denote the
probability of process (A), and the red lines denote the probability
of process (B). The temptation to defect $b=1.4$, and the average
value of rationality $T_0=1$.} \label{figWLS}
\end{figure}

\begin{figure}[t]
\centerline{\resizebox{11cm}{!}{\includegraphics{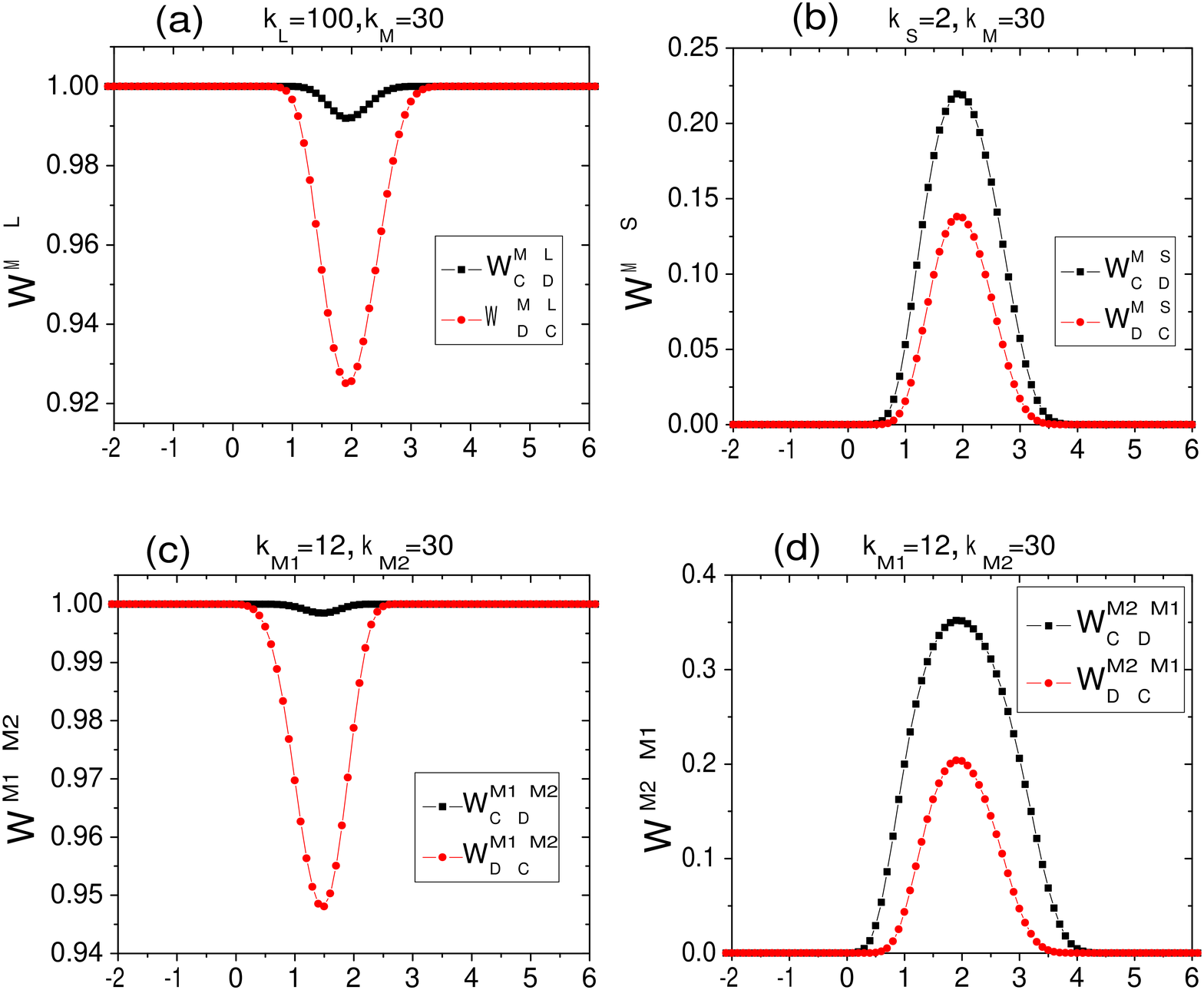}}} \caption{The
probability of middle degree nodes' strategies transiting to the
strategies of high (a), small (b), and middle (c and d) degree
nodes. In graph (a) and (c), black lines denote the probability of
process (A), and red lines denote the probability of process (B); in
graph (b) and (d), black lines denote the probability of process
(D), and red lines denote the probability of process (C). $k_L=172$,
$k_{M1}=12$, $k_{M2}=30$, and $k_S=2$. The temptation to defect
$b=1.4$, and the average value of rationality $T_0=1$.}
\label{figWZ}
\end{figure}

The strategy transition of a hub or periphery node is determined by
the four processes. For a hub node, both process (A) and (B) could
happen; for a periphery node, both process (C) and (D) could happen.
If occurrence rates of process (A) are higher than of (B), and of
(D) are higher than of (C), separately, then C clusters are
unstable, and the whole system will asymptotically be meshed in a
absorbing state of D. We can regard the strategy transition
probability in a certain process as the occurrence rate of this
process.

To calculate the four kinds of transition probability of nodes with
different degree value, we approximately classify the nodes in the
following way: (1)Nodes with small value of degree: $m\leq
k_{S}\leq\bar{k}$; (2)Nodes with middle value of degree:
$\bar{k}<k_{M}<k'$; (3)Nodes with large value of degree: $k'\leq
k_{L}\leq k_{max}$. Fig.\ref{figWLS} shows the strategy transition
probability of the high degree nodes(the upper scripts $L$, $M$, and
$S$ respectively denote nodes with large, middle, and small value of
degree). Sharp increase can be observed from the region of
\emph{cooperation crisis} (see Fig.\ref{figrouCbata}), and the
probability of process (A) is slightly higher than process (B), but
they both become equivalent to $0.5$ when $\beta$ gets larger.
However, in Fig.\ref{figWZ}, curves concerning the strategy
transition probability of the middle degree nodes present a
symmetrical fashion. More importantly, large variation of $W$ only
exists in the region of \emph{crisis}, where process (A) always
obtains larger occurrence rate than process (B), and process (D)
always obtains larger occurrence rate than process (C). These
results indicate the advantage of defectors, especially while middle
degree nodes participating in the game. In our calculation, we build
a BA network with the largest degree $k_{max}=172$ and the smallest
degree $k_{S}=m=2$. For simplicity, degrees for the three classes of
nodes are confined to isolated values. For example, here we set
$k'=50$; high degree value: $k_{L1}=172$ and $k_{L2}=100$; middle
degree value: $k_{M1}=12$ and $k_{M2}=30$; low degree value:
$k_{S}=2$; and $\mu=0.75$. Small value change brings no impact on
the qualitative results.

\begin{figure}
\centerline{\resizebox{11cm}{!}{\includegraphics{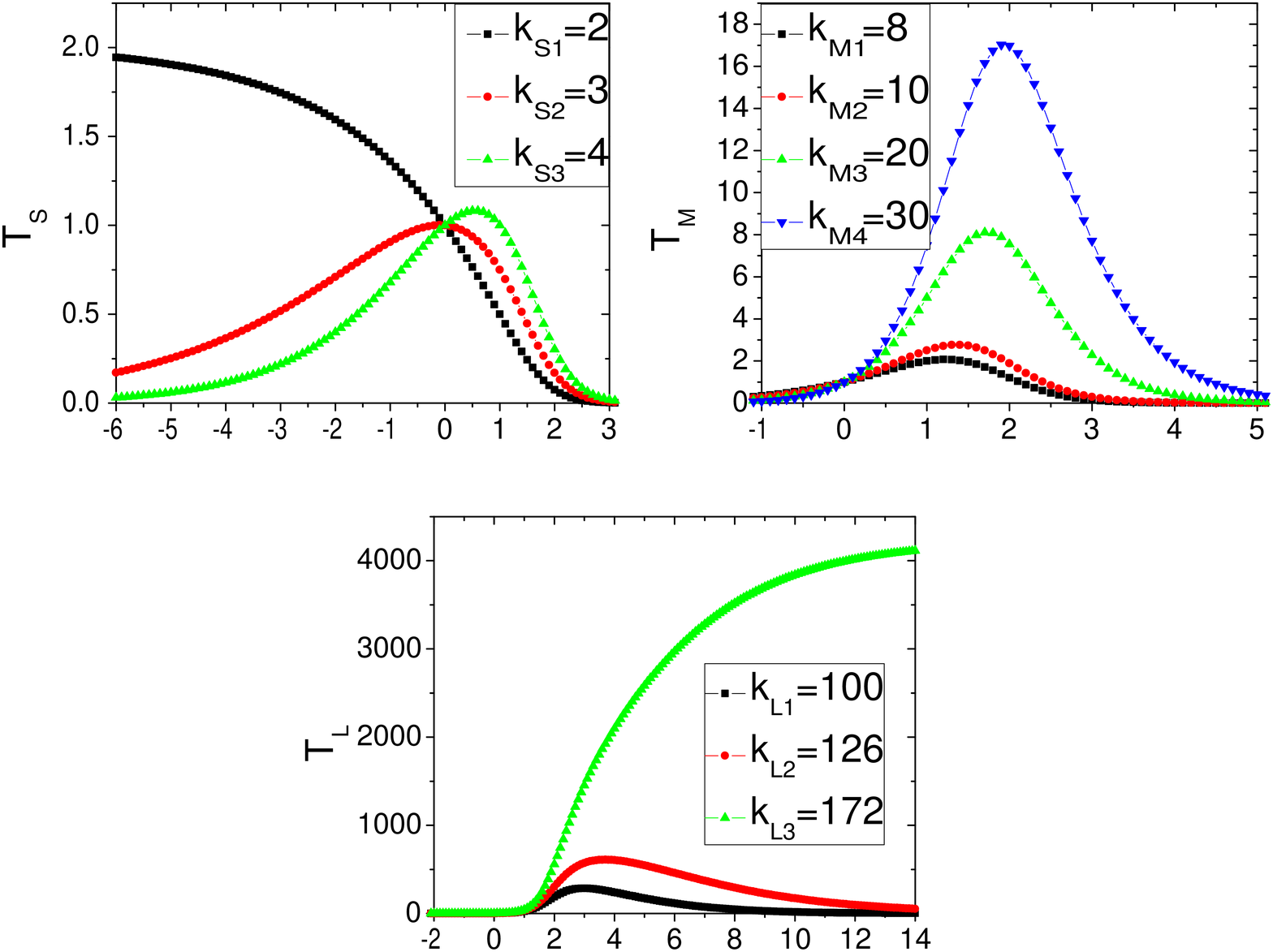}}}
\caption{The rationality of different degree nodes $T_S$, $T_M$, and
$T_L$ versus $\beta$. $k_S=2$, $3$, and $4$. $k_M=8$, $20$, and
$30$. $k_S=100$, $126$, and $172$. The average value of rationality
$T_0=1.0$.} \label{figTSML}
\end{figure}

To explore the roots of the agents' diverse behavior,
Fig.\ref{figTSML} is plotted to examine the rationality variation
versus $\beta$ for the three classes of nodes. The rationality of
the nodes with smallest and largest degree displays a monotonous
decrease and increase respectively, while that of nodes with other
degree value varies in a non-monotonous fashion. Mathematical
explanation to the numerical results is not complicated. It is
crucial to note that the peak values of $T_{M}$ are all around the
region where the \emph{cooperation crisis} takes place. These peaks
nicely explain the large variation in Fig.\ref{figWZ}. Furthermore,
the monotonous increase of rationality of the nodes with the the
largest degree results in the asymmetrical fashion in
Fig.\ref{figWLS}, as well as in Fig.\ref{figrouCbata}. Indeed,
rationality value change contributes to agents' behavior change.

In the region where the rationality of middle degree nodes reach
their peak, a node with relatively larger degree becomes rather
irrational, and thus gains a much higher probability to adopt the
strategy of a node with relatively lower payoff and lower degree.
Simultaneously, nodes with relatively lower degree, for their
irrationality, are not inclined to imitate their larger degree and
higher payoff neighbors. Consequently, this mechanism deteriorates
the validity of the cooperation-facilitating mechanism reported in
\cite{PDGscalefree,CclusterPRL2007}. As a result of the predominance
of process (A) and (D) against (B) and (C), as demonstrated in
Fig.\ref{figWZ}, defectors, though initially may be the minority in
a C cluster, do not only obtain a great chance to survive, but also
propagate fast and asymptotically dominate the whole network. This
is why the C clusters fail to maintain their stability and why
\emph{cooperation crisis} occurs. The peak rationality values of
middle degree nodes play a key role. Outside this region,
cooperative behavior is promoted.

On the left side of \emph{crisis}, especially when $\beta<0$, large
and middle degree nodes are very rational. For high and
middle-ranking defectors, when severely weakened by the low-ranking
neighbors who follow their defective strategies, low rationality
value will result in much greater sensitivity towards payoff, then a
little higher payoff of a neighbor is enough attractive for them to
imitate this neighbor, even if the degree of which is much lower.
Thereby, they gain much greater chance to transit to C strategy than
in the case of $\beta=0$, in which low-ranking players could hardly
influence the high and middle-ranking ones. Clearly, when $\beta<0$,
the efficiency of cooperation promotion is largely enhanced, even
though $T_0$ is significantly large.

On the right side of \emph{crisis}, the decision of a node with high
degree becomes random, and its strategy transition probability only
depends on the proportion between its neighboring cooperators and
defectors. This irrational hub is surrounded by large number of
rational nodes with middle and small degree, and these nodes can
quickly form a obedient domain around it and leave it with few
defective neighbors. On the other hand, highly rational neighbors of
a hub defector would first adopt the hub's strategy, simultaneously
resulting in a sudden drop of the hub's payoff, then abandon this
strategy, for the sake that the low payoff hub hardly affects them
and their cooperator neighbors with a little higher payoff could
overturn their D strategies. After that, the hub defectors will gain
much more cooperative neighbors, and thus much greater probability
to transit into a cooperator. These facts exist at high $T_0$, too.
Notably, middle degree nodes are rational outside \emph{crisis}, but
irrational in \emph{crisis}.

Based on the above discussion, as parameter $\beta$ varies from
negative value to a large positive value, the system experiences
successive sorts of dynamic processes, corresponding to the
different parts and different shapes of the curves ($\rho_c$ versus
$\beta$, show as Fig.\ref{figrouCbata}). When $\beta$ is small or
negative, $\rho_C$ is on the \emph{plateau}. For nodes with large
and middle degree, process (A), (B), (C), and (D) have approximately
the same occurrence rate, and the whole system is globally dominated
by cooperation. When $\beta$ gets larger, process (A) and (D) begin
to show considerable predominance over process (C) and (B), and
thus, the stability of C clusters is severely disturbed. When
$\beta$ is further increased, process (A) and (D) becomes
overwhelming, and C clusters are totally destroyed. $\rho_C$ decays
abruptly into the \emph{valley}, and the \emph{crisis} comes. When
$\beta$ continues to increase, the predominance of process (A) and
(D) starts to decline, and C clusters begin to resurrect. Finally,
cooperative behavior holds global prevalence, and the probability of
the four process return to a similar value. $\rho_C$ arrives at
another \emph{plateau}.

\section{Conclusion}
To sum up, in this paper, we introduce a set of rationality
distributions to investigate the effect of rationality diversity on
evolutionary prisoner's dilemma game on BA Scale-Free network. Our
model remodifies the Fermi updating rule, and our results largely
extend the results in \cite{PDGscalefree,CclusterPRL2007}. Our work
reveals that diversity of individual rationality heavily influences
the evolutionary process. Two routes, produced through two sorts of
rationality distributions (on the two sides of \emph{crisis}),
promote cooperation, even while the average value of rationality is
high. On the contrary, severe deterioration of cooperation, namely
the \emph{cooperation crisis},
also appears in another sort of rationality
distribution. By analyzing the stability of C clusters, causation of
the routes of cooperation promotion and deterioration is
interpreted.

The crucial contribution made by nodes with middle degree value may
provide some sociological inspiration. Degree is often viewed as a
certain rank of game players. Perhaps we could analogize it to some
social rank of individuals, then middle degree nodes might be
related to the middle class, which is neither the most powerful
class nor the most populous class. Middle class could serve as
social stabilizer, pointed out by Samuel P. Huntington
\cite{Huntington}. However, middle class could also display
subversive function, argued by Huntington's opponents. Individual
rationality could be affected by political or economical factors,
and are not unchangeable. Probably, since the organization of
society largely depends on the emergence of cooperation
\cite{WedekindScience2000}-\cite{OhtsukiNature2006}, such two
contrary functions could be relevant to two rationality level of
middle class in two sorts of rationality distribution. Further
investigation on the  diversity of rationality might yield new
insights towards complex social phenomenon.

\begin{acknowledgments}
This work is supported by the Natural Science Foundation of China
(No. 10775060), the Fundamental Research Fund for Physics and
Mathematic of Lanzhou University, and High Performance Computing
Center of Lanzhou University. We also thank Dr. Z.-X. Wu for helpful
discussion.
\end{acknowledgments}

\bigskip
\end{document}